# Scaling of Multi-contact Phase Change Device for Toggle Logic Operations

R. S. Khan,* N. H. Kan'an, J. Scoggin, H. Silva, and A. Gokirmak

*Department of Electrical and Computer Engineering, University of Connecticut, Storrs, Connecticut 06269, USA*

Scaling of two dimensional six-contact phase change devices that can perform toggle logic operations is analyzed through 2D electrothermal simulations with dynamic materials modeling, integrated with CMOS access circuitry. Toggle configurations are achieved through a combination of isolation of some contacts from others using amorphous regions and coupling between different regions via thermal crosstalk. Use of thermal cross-talk as a coupling mechanism in a multi-contact device in the memory layer allows implementation of analog routing and digital logic operations at a significantly lower transistor count, with the added benefit of non-volatility. Simulation results show approximately linear improvement in peak current and voltage requirements with thickness scaling.

**I. Introduction**

Phase change memory (PCM) is a state-of-the-art high-density, high-speed, high endurance non-volatile memory technology that is scalable to ~4 nm[1–5]. PCM fills the performance-capacity gap between DRAM and flash memory, with speeds comparable to DRAM and capacity comparable to flash memory. PCM utilizes the large resistivity contrast (~$10^4$ at 300 K) between the disordered amorphous (high resistivity) and ordered crystalline (low resistivity) phases of chalcogenide materials such as $Ge_2Sb_2Te_5$ (GST) to store information[6]. In recent years, there has been a growing interest in the collocation of memory and processor, as well as in-memory computation[7], to eliminate the performance bottleneck in von Neumann architectures[8], especially for machine learning applications. The dense multi-layer crossbar architecture of PCM is compatible with conventional CMOS circuitry through back-end-of-line integration[9], providing the opportunity to integrate 100s of GBs of high-speed non-volatile memory on top of the processor (FIG. 1). The CMOS realestate needed for memory access scales with the memory capacity. Hence, approaches that reduce the CMOS footprint for routing[10] and logic[11–16] functions are highly desired. Multi-contact phase change devices integrated with CMOS can significantly reduce the area requirements and provide the additional benefit of non-volatility.

The multi-contact phase change logic devices[11,14–16] utilize self-heating between pairs of contacts to electrically isolate or connect other contacts, through melting and quenching a portion ($T_{melt}$ ~873 K for GST[17]) of the phase change material or heating the material above glass transition ($T_{glass}$ ~ 420 K for GST[17]. Thermal cross-talk between the different regions of the phase change material[18] is utilized as a coupling mechanism to implement logic functions with fewer CMOS elements. Multiplexers or flip-flops can be implemented by integrating a single multi-contact device with 5 MOSFETs, while sixteen transistors are needed to implement a flip-flop and fourteen transistors are needed to implement a 2-to-1 multiplexer using conventional CMOS circuitry. Scaling of the phase change element reduces the current requirements for the MOSFETs, further reducing the CMOS realestate, improving speed and reduce power requirements. In this work, we report the results of a computational study on a 6-contact phase change device, integrated with 5 nMOSFETs, configured as a toggle flip-flop or a multiplexer (FIG. 3). Effect of scaling on power consumption and speed are reported as well.

**II. Simulation Setup**

We use 2-D finite element simulations in COMSOL Multiphysics[19] using the framework described in Woods et al.[20–22] to demonstrate the functionality and study scalability of multi-contact devices and the complementing electronic circuitry. The model simultaneously captures amorphization-crystallization dynamics, heat diffusion, electrical current flow and thermoelectric effects. Grain orientation and crystallinity are simultaneously tracked using a *crystal density* ($\overrightarrow{CD}$) vector. The components of $\overrightarrow{CD}$, $CD_1$ and $CD_2$, track grain orientation. The one-norm $||\overrightarrow{CD}|| = CD_1 + CD_2$ tracks local crystallinity: $||\overrightarrow{CD}|| = 1$ or 0 corresponds to fully crystalline or fully amorphous states, respectively. A rate equation is used to track local crystallinity:

$$\frac{dCD_i}{dt} = Nucleation + Growth + Amorphization, \quad (1)$$

where $CD_i$ is a component of $\overrightarrow{CD}$. *Nucleation* term initiates nucleation at random locations. *Growth* term is responsible for increasing $||\overrightarrow{CD}||$ to 1 once nucleation is initiated. *Amorphization* term rapidly brings $||\overrightarrow{CD}||$ to 0 at locations where temperature is greater than the melting temperature. All three terms in equation (1) are material phase and temperature

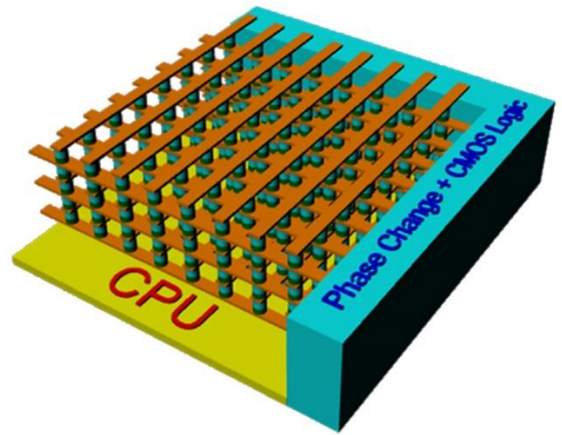

FIG. 1. Schematic of PCM crossbar array integrated on top of CPU. The phase change and CMOS logic part control routing of data and logic operations.

*Corresponding author. Email: raihan.khan@uconn.edu

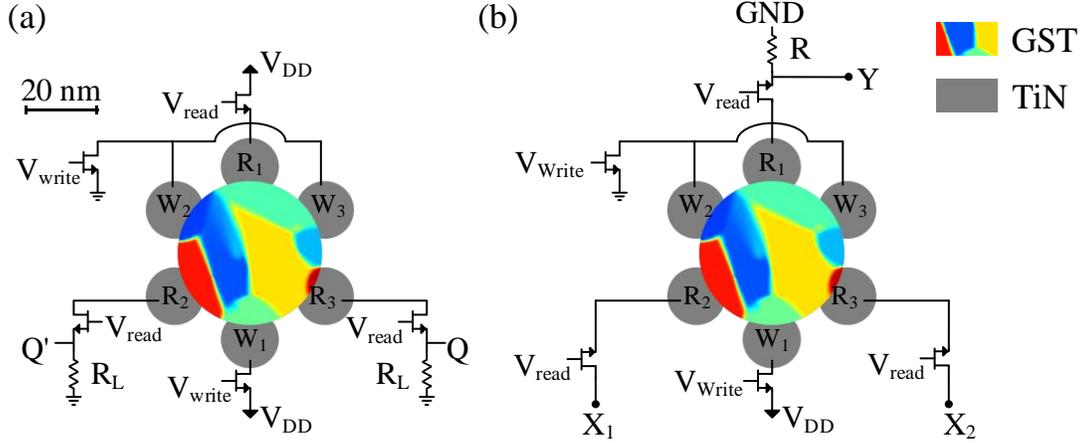

FIG. 3. Schematic of six contact toggle flip-flop (a) and toggle multiplexer (b). The distinct regions in the GST represent different crystalline grains prior to initialization of the device.

dependent. Equation (1) is coupled with electric current and heat transfer physics[23]:

$$\nabla \cdot J = \nabla \cdot (-\sigma \nabla V - \sigma S \nabla T) = 0, \quad (2)$$

$$dC_P \frac{dT}{dt} - \nabla \cdot (k \nabla T) = -\nabla V \cdot J - \nabla \cdot (JST) + Q_H, \quad (3)$$

where $J$ is the current density, $\sigma$ is the electrical conductivity, $V$ is the electric potential, $S$ is the Seebeck coefficient, $d$ is the mass density, and $k$ is the thermal conductivity. $Q_H$ accounts for latent heat of phase change $(\Delta H_{a,c})$[22].

$$Q_H = \frac{d||\overrightarrow{CD}||_1}{dt} \times \Delta H_{a,c}(T) \times d, \quad (3)$$

We use GST material parameters for the phase change material, including temperature and phase dependent electrical conductivity[24], Seebeck coefficients[24], thermal conductivities[20], and specific heats[22]. Electrical conductivity is also field dependent[25]. A constant mass density of is 6.2 g.cm$^3$ is used[26]. Although GST parameters are used here, the device concept can be implemented with other phase change materials as well. A fixed out-of-plane depth of 20 nm, 10 nm and 5 nm are used for the simulations as indicated.

### III. Device configuration

The thermal boundary conditions, and thickness and properties of the phase change material determine thermal losses, time scales for heating and cooling, hence the device speed and power requirements. The GST element is assumed to be passivated by a SiO$_2$ layer, and the thermal boundary conditons are set to be 293 K at the SiO$_2$ boundaries, located 250 nm from the center (10x times the GST radius). In the experimental phase change devices, typically a SiN layer is deposited under the oxide passivation (FIG. 2). n-channel MOSFETs with width x length = 120 nm x 22 nm are used as access devices, sized to provide sufficient current for melting and amorphization. We used TiN contacts with 10 nm radius, distributed uniformly around the circular GST patch (FIG. 3).

Three of the contacts are configured for writing (W$_1$, W$_2$, and W$_3$) and the other three for reading (R$_1$, R$_2$, and R$_3$), forming two write paths (W$_{1-2}$ and W$_{1-3}$) and two read paths (R$_{1-2}$ and R$_{1-3}$). All six contacts are accessed using nFETs. The gates of read nFETs are connected to V$_{read}$ (V$_{read}$ is high during read), and the write nFETs are connected to V$_{write}$ (V$_{write}$ is high during write). The write circuitry is identical for the flip-flop and the multiplexer implementations. Contacts W$_2$ and W$_3$ are electrically shorted and connected to ground through a single nFET. W$_1$ is connected to V$_{DD}$ (positive supply voltage) using another nFET. A short pulse at V$_{write}$ results in one of the write paths getting amorphized resulting in different device configurations. The Read circuitry is different for the two implementations: outputs Q and Q´ are taken across resistors R$_L$ connected to read contacts R$_2$ and R$_3$ for the flip-flop, whereas inputs $X_1$ and $X_2$ are applied to contacts $R_2$ and $R_3$, respectively, and the output Y is taken across resistor R connected to contact $R_1$.

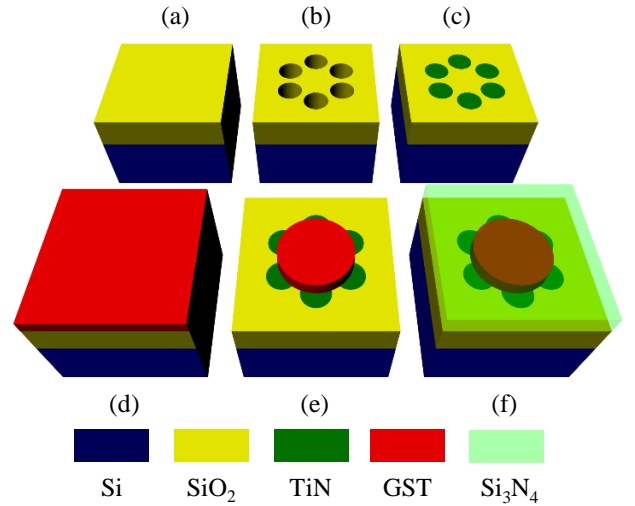

FIG. 2. Fabrication schematics for a bottom-contacted multi-contact device: (a) Growth of SiO$_2$ on Si, (b) via formation for 10 nm radius bottom contacts using optical lithography and reactive ion etching (RIE), (c) metal (TiN) deposition and planarization, (d) GST sputter deposition, (e) patterning of 25 nm radius GST patch using optical lithography and RIE, (f) deposition of Si$_3$N$_4$ for passivation.

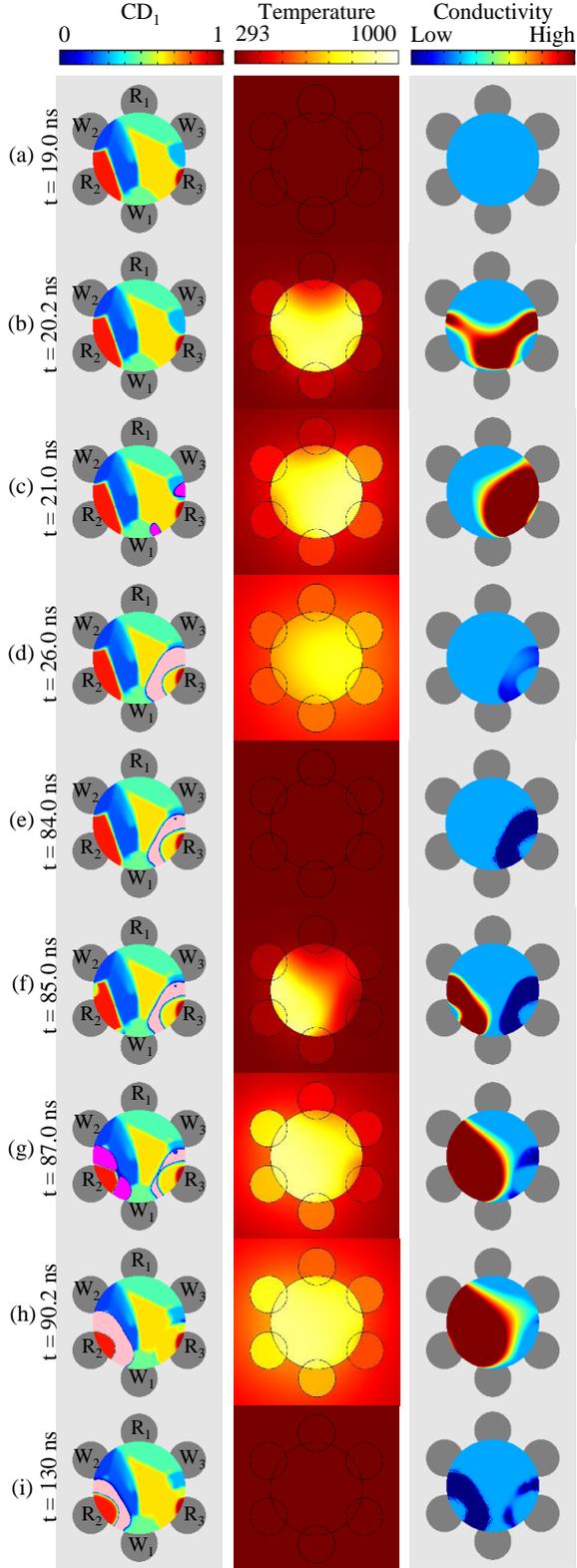

FIG. 4. Snapshots of electro-thermal simulation of the proposed device during initialization (a-d) and the first toggle operation (e-i). The $CD_1$ map (left) shows the crystallinity profile of the device at different time steps with molten and amorphous material indicated by pink and peach, respectively. The temperature map (center) shows the temperature throughout the device. The conductivity map (right) shows the conductivity of GST, where conductivity is lowest for amorphous (dark blue) and highest for melt (dark red).

## IV. Results and Discussion

We start with the device in a fully crystalline state (FIG. 4a). To initialize the device to one of the configurations, a write pulse is applied ($V_{write}$ is high). The two write paths ($W_{1-2}$ and $W_{1-3}$) begin drawing current (FIG. 5, 8 Top); one of the paths draws a progressively larger proportion of the current (FIG. 4b) and eventually melts due to thermal runaway. The 'winning' path is determined by the path resistance mismatch, which is determined by the initial random grain map and process variations. Adding a small series resistance to one of the paths deterministically set the first "winner". $W_{1-3}$ melts in this case (FIG. 4c) and amorphizes after the pulse is terminated (FIG. 4d). The device takes ~50 ns to thermalize (return to room temperature), after which a read operation is performed. The path $R_{1-3}$ is now blocked by the amorphous strip $W_{1-3}$, while $R_{1-2}$ is crystalline and thus draws much more current than $R_{1-3}$ during read (FIG. 4e). Applying a subsequent write pulse, $W_{1-2}$ draws most of the current and eventually melts because $W_{1-3}$ is initially amorphous (FIG. 4f). As $W_{1-2}$ melts, the amorphous GST in the $W_{1-3}$ path is heated above the crystallization temperature (FIG. 4g) and crystallizes: $W_{1-2}$ is now amorphous and $W_{1-3}$ is crystalline (FIG. 4h), resulting in a toggled state ($R_{1-2}$ blocked, $R_{1-3}$ crystalline). Now $R_{1-3}$ draws substantially more current that $R_{1-2}$ during read.

### A. Toggle Flip-flop

In the toggle flip flop, the output voltage toggles between low and high each time the input is high[27]. If the input is low, the output remains in its current state; if the input is high, the output switches from low to high, or from high to low. In our proposed device, the input is $V_{write}$. When $V_{write}$ is high, one of the write paths melts and amorphizes while the other crystallizes, resulting in a toggled state. As the read path that is blocked by the amorphous region draws much less current than the opposite read path, the output voltages Q and Q´ assume opposite values (when Q is high Q´ is low, and vice versa). The amorphized path and the output voltages thus toggle with each write pulse (FIG. 5). Q and Q´ can be connected to a comparator or to the gate of another transistor for rail-to-rail output. The output voltages Q and Q´ depend on the read resistor $R_L$. A higher value of $R_L$ results in higher output voltages but a reduced output ratio (FIG. 6). A higher $R_L$ reduces both $V_{GS}$ and $V_{DS}$ for the nFET. Device can be read 5ns after the write operation with a 10x contrast between the read terminals, and the contrast increases to 60x after device cools down to room temperature (FIG. 7) for the parameters used in this study, for a GST out of plane thickness of 20 nm. The ratio of the output voltages increases during cool down as the resistivity of amorphous GST decreases exponentially with temperature[24]. Thus, the cooling rate, the read resistors, and the sensitivity of the comparator all influence the speed at which the device can operate.

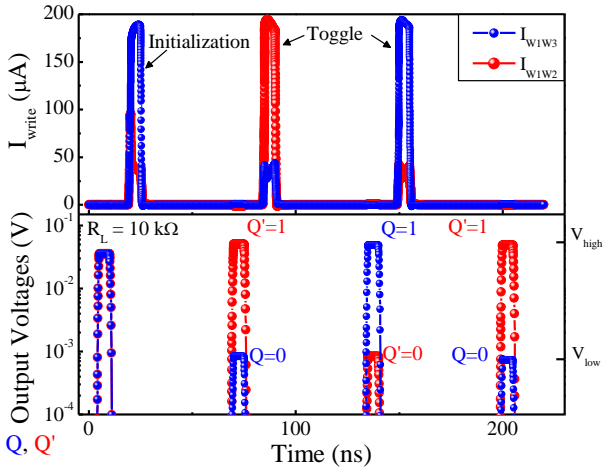

FIG. 5. Currents in the two write paths $W_{1-2}$ and $W_{2-3}$ (top) and output voltages (bottom) during the initialization pulse followed by two write pulses for toggle filp-flop. The output voltages (Q and Q´) toggle between $V_{low}$ and $V_{high}$ after each write pulse. The value of $R_L$ is 10 k$\Omega$.

## B. Toggle Multiplexers

In a multiplexer, one of several input signals are forwarded to the output based on the control pin configuration[27]. The control pin decides which input goes to the output. For the proposed toggle multiplexer, two signals ($X_1$ and $X_2$) are applied as input and there is one output (Y). $V_{write}$ is the control signal. Application of a write pulse ($V_{write}$ high) would toggle the input signal that goes to the output. With the application of the first write pulse, the the write path $W_{1-3}$ melts and amorphizes, blocking the read path $R_{1-3}$, connecting the output Y to the input $X_1$ ($X_1 \rightarrow Y$). When $X_1$ is high/low, Y follows $X_1$, irrespective of the state of $X_2$ (FIG. 8). With consequtive write pulse, $W_{1-2}$ is amorphized, blocking read path $R_{1-2}$, switching Y to $X_2$ ($X_2 \rightarrow Y$), irrespective of state of $X_1$. Thus each write pulse toggles the input signal that is forwarded to the output.

## C. Discussion

As the device thickness is scaled down, the GST volume required to be amorphized decreases. This results in decrease in energy and power consumption. This is verified by simulations as maximum power consumption for the device decreases from 585 µW to 92 µW as thickness is scaled from 20 nm down to 5 nm. There is reduction in the supply voltage $V_{DD}$, write current $I_{write}$ (FIG. 9), and energy consumption with scsaling. Parameters for different thicknesses are compared in Table I.

The speed of the device is determined by the distance between write contacts (shorter distance between contacts will result in faster re-crystallization of amorphous paths, thus higher speed), placement of thermal anchor (the closer the thermal anchors, the less time it will take to cool down to room temperature, though at the cost of additional power for write operations), and the size of the write FETs (larger FETs provide more current at the cost of an increased footprint). Although we simulated the device concept with a 25 nm radius GST patch, the device can be scaled even further, limited by the fabrication of the TiN contact and the heat transfer within the GST patch that must allow for sufficient but not too much thermal cross-talk. A smaller GST patch would reduce the power consumption and required FET size as well, or increase the speed of the device for the same FET size, at a reliability cost in terms of thermal cross-talk control. For example, the minimum time required to successfully amorphized a write path decreases from 6 ns to 5 ns as GST radius is decreased from 35 nm to 25 nm.

For both the flip-flop and the multiplexer the $I_{write}$ through the path that is recrystallizing ($I_{W1W3}$ for the first write pulse after initialization and $I_{W1W2}$ for second write pulse) increases during the pulse duration. This is the combined effect of decrease in amorphous resistivity and recrystallization of previously amorphous write path due to increase in temperature. With the resistivity of the path decreasing, it starts drawing current. If the pulse is not terminated in due time, both paths will eventually melt, resulting in device failure (The device can be returned to operational state by annealing/set pulse). Thus careful consideration of pulse duration is needed to ensure proper operation of the device. We did not observe any degradation in the simulated output voltages with repeated write pulses. In experimental devices void formation at write contacts during higher cycle counts may be a reliability concern.

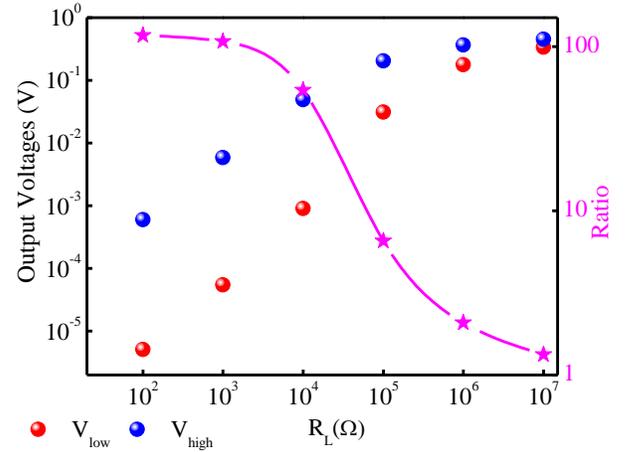

FIG. 6. Output voltages for different values of $R_L$ for the toggle flip-flop (spheres, left-axis). Low and high values are represented using red and blue spheres respectively. The right axis (stars) shows the ratio of $V_{high}$ to $V_{low}$.

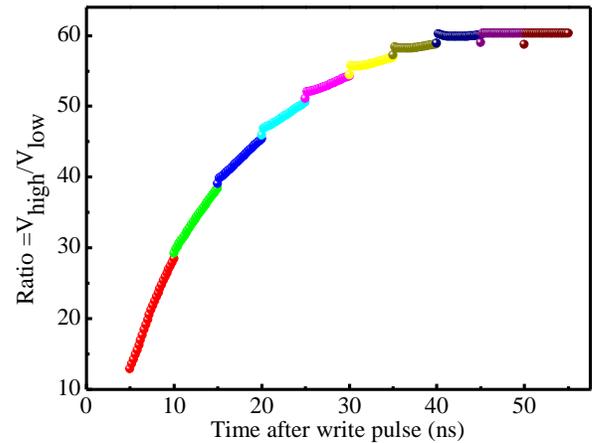

FIG. 7. Ratio of $V_{high}$ to $V_{low}$ for the toggle flip-flop during 5 ns read pulses after amorphization. The first read is performed 5 ns after the termination of the write pulse. The change in ratio during the read pulse is due to change in temperature of the device. The ratio stabilizes after 40 ns following the termination of the write pulse.

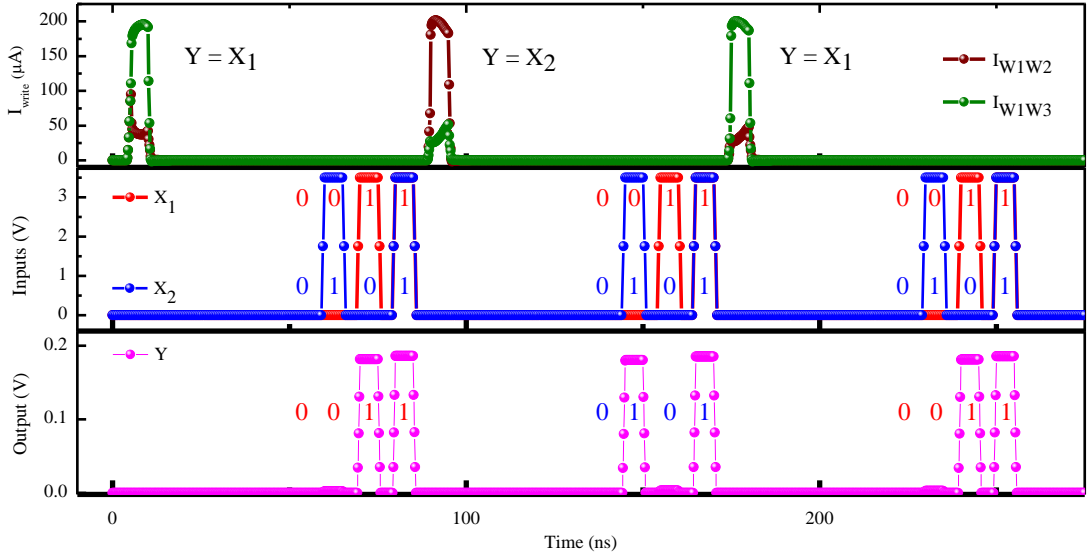

FIG. 8. Write currents (top), input voltages (middle), and output voltages (bottom) for toggle multiplexer. For read operations, four combinations of $X_1$ and $X_2$ are tested $(X_1, X_2)$: (0, 0), (0, 1), (1, 0), (1, 1). Output follows the state of one of the inputs (determined by the write pulse) regardless of the state of the other input.

Even in ideal circumstances, the proposed device will be slower than its CMOS implementation: ~.3 ns for CMOS writes vs. ~10 ns for the proposed device (FIG. 7, assuming the comparator has the sensitivity to detect a ratio of ~30x between output voltages). The presented configurations in FIG. 3 do not provide rail-to-rail functionality, which can be addressed by altering the access circuitry or the device configuration. For example, using an inverter instead of a resistor in the read circuit for the toggle multiplxer can achieve rail-to-rail voltage (FIG. 10). A conventional CMOS toggle flip flop and 2-to-1 multiplexer requires 16 and 14 transistors, respectively. The presented device uses 5 transistors to achieve similar functionality, thus reducing the FET count significantly and reduce the necessary CMOS realestate by ~2x.

### V. Summary

A computational analysis of six-contact device with toggle flip-flop or multiplexing functionality using only 5 transistors (compared to 16 and 14 in a conventional flip-flop and multiplexer) and non-volatility has been presented. 2-D finite element simulations were performed using temperature dependent material parameters and accounting for thermoelectric effects. The pulse duration and required transistor sizes depend on device dimensions, contact spacing, and the phase change material. The presented approach to integrate a phase change element with CMOS bring additional functionality (e.g. logic, routing) to the memory layer to free up precious CPU resources and require no power dissipation to hold information hence are also suitable for intermittent power applications.

| Table I- Write/read pulse parameters | | | |
|---|---|---|---|
| | Out of plane thickness (nm) | | |
| | 20 | 10 | 5 |
| $V_{DD}$ (V) | 3 | 2.2 | 1.65 |
| Peak write voltage (V) | 3 | 2.2 | 1.65 |
| Rise /fall time for read and write pulses (ns) | 1 | 1 | 1 |
| Read/write pulse duration (ns) | 5 | 5 | 5 |
| $V_{read}$ during read operation (V) | 0.5 | 0.5 | 0.5 |
| Peak write current (µA) | ~193 | 101 | 56 |
| Peak read current (µA) | 6 | ~4.7 | ~4.2 |
| Maximum power (µW) | 585 | 222 | 92 |
| Write energy (pJ) | ~2.9 | ~1.1 | ~0.46 |

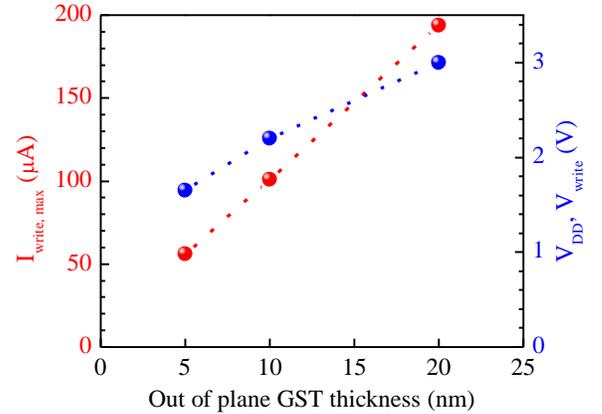

FIG. 9. Peak current during write operation (red spheres, left axis) and supply voltage (blue spheres, right axis) for different out of plane GST thicknesses. Both the current and the voltage decrease with decreasing thickness.

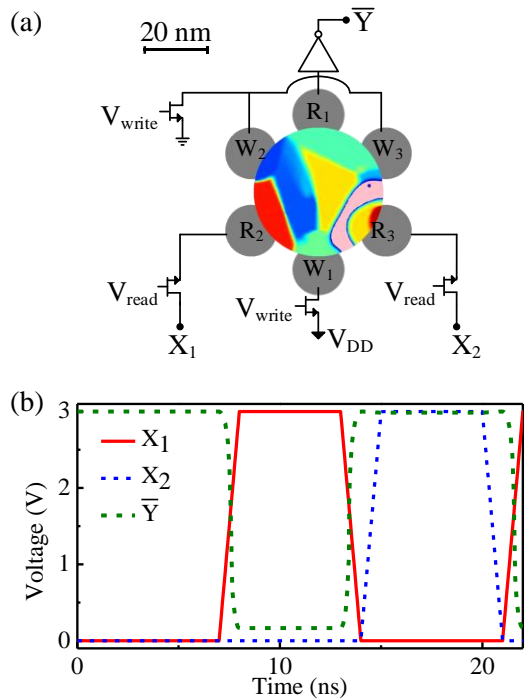

FIG. 10. Toggle multiplexer with modified read circuitry for rail-to-rail output voltage (a). Complement of output voltage Y for different $X_1$ and $X_2$ values. Simulation is performed in LTspice.

**Acknowledgements**

Raihan Khan performed the simulations, analysis and writing of the manuscript supported by the U.S. National Science Foundation (NSF) through ECCS 1711626 award. Nadim H. Kan'an formulated the idea and performed preliminary simulations supported by NSF ECCS CAREER 1150960 award. Ali Gokirmak and Helena Silva contributed to the design of experiments, analysis and writing of the manuscript.

**Data Availability**

The data that support the findings of this study are available from the corresponding author upon reasonable request.